\documentclass[aps,pre,twocolumn,groupedaddress,showpacs]{revtex4}
\usepackage{amsmath}
\begin{document}

\title{Gaussian noise and time-reversal symmetry in non-equilibrium Langevin models}

\author{M. H. Vainstein}
\affiliation{Instituto de F\'{\i}sica and Centro Internacional de
F\'{\i}sica da Mat\'eria Condensada, Universidade de Bras\'{\i}lia, CP 04513, 70919-970, Bras\'{\i}lia-DF, Brazil}
\author{J. M. Rub\'{\i}}
\email[corresponding author: ]{mrubi@ub.edu}
\affiliation{Departament de F\'{\i}sica Fonamental, Universitat de Barcelona, Diagonal 647, 08028 Barcelona, Spain}

\date{\today}

\begin{abstract}
We show that in driven systems the Gaussian nature of the fluctuating force and time-reversibility are equivalent properties. This result together with the potential condition of the external force drastically restricts the form of the probability distribution function, which can be shown to satisfy time-independent relations. We have corroborated this feature by explicitly analyzing a model for the stretching of a polymer and a model for a suspension of non-interacting Brownian particles in steady flow.
\end{abstract}


\maketitle
\section{Introduction}
The advances of experimental techniques have permitted the manipulation of single molecules and the study of their behavior under magnetic~\cite{Smith92} and hydrodynamic forces~\cite{Perkins95}.  These experiments have made it possible to analyze processes taking place at very short length scales and have opened new perspectives on applications of non-equilibrium statistical mechanics to small-scale systems. Although being under non-equilibrium conditions, many of these systems obey  time-independent relations, or generalized fluctuation-dissipation theorems, whose importance has been stressed in a number of  experimental situations~\cite{Wang02,Carberry04,Shang05,Douarche06}.
 It has been shown~\cite{Astumian06} that these relations can be derived from the Onsager-Machlup theory~\cite{Onsager53}, which  assumes that the noise in the Langevin equation is Gaussian. In this work, we find that the relations are a direct consequence of the time-reversal symmetry principle. We show a theorem connecting both properties when the system is subjected to an external driving force. Our conclusion is that time-reversibility has important consequences on the stochastic behavior of non-equilibrium systems, leading in particular to restrictions on the probability distribution.  

The paper is organized as follows: in Sec.~\ref{sec.langevin}, we study Langevin equations in one dimension which have been used as a model for the stretching of single molecules and for the dynamics of colloidal particles in a translating optical trap. We also analyze the implications of time-reversal symmetries in such systems.  In Sec.~\ref{sec.lin_flow} we consider the Brownian motion of non-interacting particles in linear two-dimensional stationary velocity fields, and compare the results with those of the one-dimensional case. In Sec.~\ref{sec.nonlin_flow} we briefly discuss Brownian motion in the nonlinear Poiseuille flow, and give concluding remarks in Sec.~\ref{sec.conclusion}.

\section{Langevin models and detailed balance} 
\label{sec.langevin}

Many methods, such as scanning probe microscopy and laser optical traps, have been deviced to study the mechanical properties of  DNA, modular proteins or synthetic polymers by the extension of a single molecule~\cite{Wang01}. In former of these methods, the tip of the atomic force microscope  is subject to fluctuations.  In the latter,  the experiments on the stretching of single molecules are usually done by attaching a polymer to two beads, one of which is kept fixed and the other moved by an optical tweezer. The moving bead is subject to thermal fluctuations due to the solvent medium, which is large enough to be undisturbed and considered as a thermal bath at a fixed temperature $T$.  As such, the extension of the molecule can be modelled by an overdamped Langevin equation for the position of the bead
 
\begin{equation}
\label{eq.langevin}
\gamma\dot{x}=-V'(x,t)+\xi(t),
\end{equation}
where $\gamma$ is the friction coefficient of the particle, $\xi(t)$ is an additive random force, $V(x,t)$ is a potential and the prime denotes a space derivative. It is usual to consider the overdamped case since the low Reynolds number regime accounts for many situations, such as the motion of microorganisms and macromolecules in solution~\cite{Astumian06,Purcell77}. For the potential, we take 
\begin{equation}
\label{eq.potential}
V(x,t)=V_0(x)-H(t-t_0)Fx,
\end{equation}
 where $H(t)$ is the Heaviside step function, to account both for the internal potential of the molecule $V_0(x)$ and for a constant external driving force $F$, turned on at time $t_0$. Since the medium is in thermal equilibrium, it is common to assume that the random force is a Gaussian white noise with zero mean which obeys the fluctuation-dissipation relation 
\begin{equation}
\langle \xi(t) \xi(t')  \rangle = 2\gamma k_B T \delta(t-t'),
\end{equation}
 where $k_B$ is Boltzmann's constant.

 A driven system such as the dragged Brownian particle considered experimentally by Wang \emph{et al.}~\cite{Wang02} can be modelled by Eq.~(\ref{eq.langevin}). In the experiment, an optical trap with a harmonic potential near the focal point is translated relative to the solvent with constant velocity $v_{opt}$. The optical force acting on the colloidal particle at position $x$ is given by $F_{opt}=-k(x-x_0)$, where $x_0$ is the center of the trap. Therefore, in the laboratory coordinates, the system can be modelled by
\begin{equation}
\gamma\dot{x}=-k(x-v_{opt}t)+\xi(t).
\end{equation}
By making a change of coordinates to the comoving frame $x_1=x-v_{opt}t$, the equation becomes
\begin{equation}
\gamma\dot{x}_1=-\gamma v_{opt}-kx_1+\xi(t),
\end{equation}
which has precisely the form of Eq.~(\ref{eq.langevin}) with a potential of the form of Eq.~(\ref{eq.potential}). This system has also been  studied in the formalism of the Onsager-Machlup theory~\cite{Astumian06a,Taniguchi06}.

The Gaussian nature of the stochastic force has implications on the behavior of the system under time-reversal. In this case, it is possible to use the Fokker-Planck equation to describe the evolution of the probability distribution $P(x,t)$ associated to the stochastic processes given in Eq.~(\ref{eq.langevin})
\begin{align}
\label{eq.FP} 
\frac{\partial }{\partial t}P(x,t) &=\frac{\partial }{\partial x}\left\{ \frac{1}{\gamma}V'(x,t) P(x,t)+ D \frac{\partial }{\partial x}P(x,t) \right\} \nonumber\\
&\equiv \hat{L}_{FP}(x,t)P(x,t),
\end{align}
 where $D\, =\, k_B T/\gamma$, and  $\hat{L}_{FP}(x,t)$ is the Fokker-Planck operator. The stationary distribution is given by  
\begin{equation}
P_s(x,t)=Z^{-1}e^{-V(x,t)/k_B T},
\end{equation}
  where $Z$ is the normalization constant and the dependence on $t$ comes from the step function and indicates only that there are two stationary states, depending on whether the force $F$ is present or not. To simplify the notation, we will drop the $t$ dependence from now on. 
 The Fokker-Planck equation  implies time-reversibility~\cite{Risken89,Qian02}, which is given by the operator relation 
\begin{equation}
\hat{L}_{FP}(x) P_s(x)=P_s(\tau x)\hat{L}^{\dagger}_{FP}(\tau x),
\end{equation}
 where  $\hat{L}^{\dagger}_{FP}(x)$ is the adjoint operator  and time-reversal is indicated by $\tau=1$ ($\tau=-1$) if $x$ is an even (odd) variable.

On the other hand, it can be shown  that time-reversal on the microscopic level implies that the random force is Gaussian.  In general, the probability distribution $P(x,t)$  evolves  following the master equation
\begin{equation}
\frac{\partial P(x,t)}{\partial t}=\int dx' [\hat{T}(x|x') P(x',t)-\hat{T}(x'|x) P(x,t)],
\end{equation}
which can be written in the form of the Kramers-Moyal expansion~\cite{vanKampen92}
\begin{equation}
\frac{\partial P(x,t)}{\partial t}=\sum_{n=1}\frac{(-1)^n}{n!} \left( \frac{\partial}{\partial x}\right)^n  \{ a_n(x)P\} ,
\end{equation}
where $a_n(x)$ are the moments of the transition probabilities $\hat{T}(x|x')$, which can be shown,  by using  the additive nature of the noise,  to be given by 
\begin{equation}
\hat{T}(x|x')=\hat{T}_0(x|x')\delta(x-x'),
\end{equation}
 where  
\begin{equation}
\label{trans_prob1}
 \hat{T}_0(x|x')\!=\!\left[ \frac{1}{\gamma} \frac{\partial }{\partial x}V'(x) + \sum_{n=2}^{\infty} \frac{(-1)^n}{n!}\alpha _n  \frac{\partial^n }{\partial x^n}\right].
\end{equation}
Here, the $\alpha_n$'s are the  cumulants of the stochastic force $\gamma^{-1}\xi(t)$, thus establishing a connection between a Langevin model and the Kramers-Moyal expansion. 
 It is also possible to express the force $V'(x)$ in terms of the stationary distribution as 
\begin{equation}
\label{force}
\frac{1}{\gamma}V'(x)=-\sum_{n=2}^{\infty} \frac{(-1)^n}{n!}\alpha_n \left[ \frac{1}{P_s(x)}\frac{\partial^{n-1}}{\partial x^{n-1}}  P_s(x) \right],
\end{equation}
 which in the case of Gaussian noise simplifies to
\begin{equation}
\frac{1}{\gamma}V'(x)=- \frac{\alpha_2}{2}\frac{\partial}{\partial x} \ln P_s(x),
\end{equation}
  the definition of a thermodynamic force~\cite{deGroot84}.
 
 The condition of  microscopic time-reversibility is given by 
\begin{equation}
\hat{T}(x'|x)P_s(x)=\hat{T}(\tau x|\tau x')P_s(x'),
\end{equation} 
 where  $P_s(x)=P_s(\tau x)$. From here, it can be easily shown that the moments 
\begin{equation} 
\label{moments}
T_{lm}\equiv \int dx dx' x^l \hat{T}(x|x')P_s(x')(x')^m
\end{equation} 
obey the symmetry relation~\cite{Mazur91}
\begin{equation}
\label{symmetry}
 T_{lm}=\tau^{l+m}T_{ml}.
\end{equation}
On substituting Eqs.~(\ref{trans_prob1}) and (\ref{force}) into Eq.~(\ref{moments}) and proceeding by integration by parts, one arrives at
\begin{align}
T_{ml}=-l\sum_{n=2}^{l+m} \frac{1}{n} \alpha_n &\left[ \binom{l+m-1}{n-1} \right. \nonumber \\
&\left.  -\binom{l-1}{n-1}\theta_{l-n}\right] \langle x^{l+m-n}\rangle,
\end{align}
where $\theta_j \equiv 1$ if $j\geq 0$ and $\theta_j \equiv 0$ if $j < 0$. By applying relation (\ref{symmetry}) to the moment with $l=1$ and $m=2$, one finds that $\alpha_3=2\tau \alpha_3$, and consequently $\alpha_3=0$. Then, assuming that $\alpha _n=0$ up to $j-1$ and considering the moments with $l=1$ and $m=j-1$, one obtains 
\begin{equation}
\alpha_j =(j-1)\tau^j \alpha_j,
\end{equation}
  to conclude by induction that $\alpha_n=0$, for $n>2$ (for the case in the absence of an external force, see Ref.~\cite{Mazur91}). Therefore, time-reversal implies that the random force is Gaussian and that the Fokker-Planck equation is also valid in this case.

The probability of observing a particle moving from $x_1$ at time $t_1$ to $x_n$ at time $t_n$ by any trajectory must be the same as the probability of observing  the inverse trajectory in the equilibrium state
\begin{equation}
\label{eq.trans}
P_s(x_n)P(x_1,t_1|\cdots|x_n,t_n)=P_s(x_1)P(x_n,t_n|\cdots|x_1,t_1),
\end{equation}
where $P(x_1,t_1|\cdots|x_n,t_n)$ represents the joint probability of being in the positions $x_j$ at time $t_j$.  
 From here, we then arrive at the detailed balance condition
\begin{equation}
\label{eq.transition1}
\frac{P(x_1,t_1|\cdots|x_n,t_n)}{P(x_n,t_n|\cdots|x_1,t_1)}=e^ {- [V(x_1)-V(x_n)]/k_B T},
\end{equation}
which, despite being derived using the stationary probability,  is valid even if the system  has not yet reached a stationary state, because it is a necessary condition on the dynamics of the system to guarantee that it reaches stationarity with the correct Boltzmann weights.

The important aspect is that one is dealing with the dynamics of a potential system in which the transition probabilities depend only on energy differences. This can be seen from the fact that the Fokker-Planck operator of potential systems can be cast in the form
\begin{equation}
\label{potential_FP}
\hat{L}_{FP}(x)=D\frac{\partial}{\partial x}e^{-V(x)/k_B T}\frac{\partial}{\partial x}e^{V(x)/k_B T}.
\end{equation}

Before the external force $F$ is switched on, the bead is in equilibrium fluctuating around the minimum of $V_0(x)$ and satisfies relation~(\ref{eq.transition1}).
 At the moment the force is turned on, the position of the minimum of the potential changes, the bead will no longer be at equilibrium and will take some time to relax to its new state of equilibrium with a force~\cite{terHaar60} around the new minimum. However, relation~(\ref{eq.transition1}) continues being valid, but with the new potential,  since it should be valid independently of the system being in an equilibrium or a non-equilibrium state.

That this should be the case is not unexpected, since from the point of view of the trapped bead undergoing Brownian motion, the origin of the potential force is irrelevant. The particle always has a small probability of climbing the potential since its thermal motion never ceases and does not change by the switching on of the external force. From our point of view, it may appear strange that in pulling the particle in one direction it can move in the other; however, it is a natural consequence of the detailed balance relation. 
 
In the one-dimensional case studied thus far, there is always a potential if the force is a function of position only, which is not so for higher dimensions, when non-potential contributions may exist. 
 
\section{Brownian motion in linear flows}
\label{sec.lin_flow}
To illustrate the implication of the lack of a potential, we consider the motion of particles in an infinite incompressible liquid in stationary flow, taken to be small enough and to move at low velocities so as to not perturb significantly the velocity field (small P\'eclet number regime). In the presence of large friction, the inertial effects can be neglected.  The probability distribution $P(\mathbf{r};t)$ of spherical point-like colloidal particles in the liquid with velocity field $\mathbf{v}(\mathbf{r})$ can be calculated from the Smoluchowski equation~\cite{Risken89}
\begin{equation}
\label{eq.smol}
\frac{\partial P}{\partial t}+(\mathbf{v} \cdot \nabla)P = D\nabla^2 P.
\end{equation}
As noted earlier, in writing this equation, the assumption of Gaussian noise in the Langevin formalism is implicitly made. The  corresponding equation is similar to Eq.~(\ref{eq.langevin}), except that in this case it is a vector equation with the force term $-V'(x,t)$ replaced by the stationary velocity field $\mathbf{v}(\mathbf{r})$. In this work we will address the class of planar stationary flows whose linear velocity field  has components 
\begin{equation}
v_x=Gy \quad \text{and} \quad v_y=\alpha Gx, \nonumber
\end{equation}
 where $G$ is a constant shear rate and  $\alpha$ is a parameter that can range from $-1$ (pure rotation), through zero (simple shear) to $1$ (pure elongation). In this case, it is possible to calculate analytically~\cite{van_de_Ven80}  the distribution function $P(\mathbf{r};t)$, for  the initial and boundary conditions
\begin{equation}
 P(\mathbf{r},0)=\delta(\mathbf{r}),\quad \text{and}\quad \lim_{\mathbf{r}\to \infty}P(\mathbf{r},t)=0,
\end{equation}
 respectively. The solution is the conditional probability that a particle initially at the origin is at position $\mathbf{r}$ at time $t$, and is given by the generalized Gaussian
\begin{eqnarray}
\label{eq.prob_distr}
P(\mathbf{r},t)=\frac{\alpha}{2\pi}\left( \frac{2G}{D\chi(t)}\right)^{\frac{1}{2}} \exp \left\{ -\frac{\alpha^{\frac{1}{2}}}{2\chi(t)}\left[ \alpha\psi_{+}(t)x^2 \right. \right.  \nonumber \\  
\left. \left.+ \psi_{-}(t)y^2 -2\alpha^{\frac{1}{2}}(\alpha+1)\phi(t)xy  \right] \right\}, 
\end{eqnarray}
 where
\begin{align}
\psi_{\pm}(t)&=(\alpha+1)\sinh(2\alpha^{\frac{1}{2}}Gt)\pm 2\alpha^{\frac{1}{2}}(1-\alpha)Gt, \nonumber\\
\chi(t)&=DG^{-1}\{(\alpha+1)^2[\cosh(2\alpha^{\frac{1}{2}}Gt)-1] \nonumber \\ &\quad  -2\alpha(\alpha-1)^2G^2t^2 \}, \nonumber \\
\phi(t)&=\cosh(2\alpha^{\frac{1}{2}}Gt)-1. \nonumber
\end{align}
It is interesting to notice that even in the presence of large fluxes the solution is Gaussian. Therefore, this is an example of a system which can be driven far from equilibrium and still be characterized by Gaussian distributions. The probability distribution  obtained for the infinite system does not have a stationary solution, as can be seen by taking the limit $t\to \infty$ in Eq.~(\ref{eq.prob_distr}), which leads to 
\begin{equation}
\lim_{t\to \infty}P(\mathbf{r},t)=0,
\end{equation}
 for all $\mathbf{r}$. This is expected, since the velocity field considered allows the particles to spread out without bound. In order to generate a stationary distribution, it would be necessary to confine the system to a finite region by using reflective walls and/or to consider periodic boundary conditions (as in Couette flows between two cylinders). We show that even though the generalized Onsager relations~\cite{PhysRevA.36.222} are satisfied by the mobility matrix of a Brownian particle in all linear flows~\cite{Rubi91},  the time-independent relations  are valid only when the velocity field is potential.

\subsection{Elongational flow}
 For elongational flow ($\alpha=1$), the flux is potential and it is possible to write $\mathbf{v}=-\nabla \Phi$, where
\begin{equation}
\Phi(x,y)=-Gxy. 
\end{equation}
 The probability distribution function is given by 
\begin{align}
&P(\mathbf{r};t)=\frac{|G|}{2^{\frac{3}{2}}\pi D[\cosh(2Gt)-1]^{\frac{1}{2}}} \nonumber \\
&\times \exp \left \{ \frac{-G}{4D}\left[\frac{\sinh(2Gt)}{\cosh(2Gt)-1} \right](x^2+y^2) + \frac{G}{2D}xy \right\},
\end{align}
which displays the time-independent relations 
\begin{equation}
\label{eq.ft1}
\frac{P(x,y;t|G)}{P(-x,y;t|G)}=\frac{P(\Phi)}{P(-\Phi)}=e^{Gxy/D}=e^{-\gamma\Phi/k_BT},
\end{equation}
together with all other similar relations obtained by transformations on $x$, $y$ and $G$ which maintain $\Phi$ invariant. These relations have the form of what has been named generalized fluctuation-dissipation theorems, since $\gamma \Phi$ is the work done on  a particle. It should be pointed out that this relation can be deduced from the detailed balance principle~(\ref{eq.transition1}) simply by assuming that the transition probabilities depend only on the energy difference between the initial and final states, what is consistent with Eq.~(\ref{potential_FP}). The reasoning goes as follows: if the system is to have the Boltzmann weights in a stationary state, then detailed balance must be satisfied. Invoking the Markov property, a transition probability  can be written as a product 
\begin{equation}
P(x_a,t_a|x_0,t_0|x_b,t_b)=P(x_a,t_a|x_0,t_0)P(x_0,t_0|x_b,t_b).
\end{equation}
Now, if the transition probabilities depend only on the energy difference, for the particular case in which $E_a=E_0+\Delta E$ and $E_b=E_0-\Delta E$, we have
\begin{equation}
\frac{P(x_a,t_a|x_0,t_0|x_b,t_b)}{P(x_b,t_b|x_0,t_0|x_a,t_a)}=
\left[ \frac{P(x_0,t_0|x_b,t_b)}{P(x_0,t_0|x_a,t_a)}\right]^2,
\end{equation}
and Eq.~(\ref{eq.ft1}) is deduced directly from here by using Eq.~(\ref{eq.transition1}).  It can be written as
\begin{equation}
\frac{P(\Delta E)}{P(-\Delta E)}=e^{-\Delta E/k_BT}.
\end{equation}
It should be noted that in performing Monte Carlo simulations with the Metropolis algorithm, one of the requirements for a system to be able to reach the correct stationary distribution is precisely this relation (the other requirement is that the algorithm be ergodic)~\cite{Metropolis53}.

From the fact that the distribution function is normalized 
\begin{equation}
\int_{-\infty}^{\infty}\int_{-\infty}^{\infty}  P(x,y;t) \,dx dy =1,
\end{equation}
 and from Eq.~(\ref{eq.ft1}) we arrive at
\begin{equation}
\label{eq.ft2}
\langle e^{-\gamma\Phi/k_BT} \rangle =\int_{-\infty}^{\infty}\int_{-\infty}^{\infty} e^{-\gamma\Phi/k_BT}P(x,y;t)\,dxdy=1
\end{equation}
This last result is valid throughout the evolution of the system and in the stationary state (when there is one). Equations (\ref{eq.ft1}) and (\ref{eq.ft2}) are analogous to the the ones obtained by Astumian~\cite{Astumian06} for a single particle in a colloidal suspension in equilibrium in the gravitational field~\cite{Chandrasekhar43} and to relations known as generalized fluctuation-dissipation theorems~\cite{Bochkov81}.

\subsection{Shear flow}
The case of a shear flow ($\alpha = 0$) shows different features. Its probability distribution is given by
\begin{align}
P(\mathbf{r};t)&=(4\pi Dt)^{-1}\left(\frac{3}{(Gt)^2+12} \right)^{\frac{1}{2}} \nonumber \\ 
&\times \exp \left\{ -\frac{3[x-\frac{1}{2}yGt]^2}{Dt[(Gt)^2+12]} -\frac{y^2}{4Dt} \right\}.
\end{align}
In this case, the ratio between the probabilities of symmetric trajectories is no longer time-independent: 
\begin{equation}
\label{eq.shear}	
\frac{P(x,y;t|G)}{P(-x,y;t|G)}=\exp\left\{\frac{6Gxy}{D[(Gt)^2+12]}  \right\}.
\end{equation}
Here, we notice the appearance of the potential of the elongational flow divided by a function of time in the exponential. This can be understood by the fact that simple shear can be viewed as a composition of a rotation, whose vorticity destroys the potential nature of the flow, and an elongational flow. For small times or small shear rates, if we expand the exponent to first order, this case will approximately obey the law found for elongational flow. However, as time passes, convection becomes important and the flow is no longer approximately potential. For large times, we can no longer make the time-independent approximation and the time-independent relations will not be valid. In this case, we cannot make the same reasoning as in the former, which is due to the fact that no potential exists for simple shear. We can consider as an approximate experimental realization of simple shear the flow of fluid between two rotating coaxial cylinders. Since the flow does not contribute a potential energy, it is expected that at long times the distribution of Brownian particles $P_s(x,y)$ should become homogeneous. We should obtain from Eq.~(\ref{eq.trans})
\begin{equation}
\label{eq.det_bal_shear}
\frac{P(\mathbf{r}_1,t_1|\cdots|\mathbf{r}_n,t_n)}{P(\mathbf{r}_n,t_n|\cdots|\mathbf{r}_1,t_1)}=1.
\end{equation}
Instead, we obtain Eq.~(\ref{eq.shear}). 
However, in taking the limit $t\to \infty$, we arrive at Eq.~(\ref{eq.det_bal_shear}), so that at long times we recover the expected stationary result. 

\subsection{Pure rotation}

The case of pure rotation ($\alpha=-1$) has a distribution function identical to when there is no external velocity field 
\begin{equation}
P(\mathbf{r},t)=\frac{1}{4\pi D t}\exp \left[ -\frac{x^2+y^2}{4Dt}\right],
\end{equation}
  which obeys Eq.~(\ref{eq.det_bal_shear}), as it should, since there is no potential involved.

\section{Brownian motion in Poiseuille flow}
\label{sec.nonlin_flow}
Up to now, we have been dealing with linear flows. We will now focus on  a nonlinear non-potential flow, which in a fixed coordinate system $(x',y')$ is given by the parabolic velocity profile 
\begin{align}
v_x'&=\gamma (R_0^2-y^2), \text{and} \nonumber \\
v_y'&=0 \nonumber,
\end{align}
 where $\gamma=V_{max}/R_0^2$, $R_0$ is the tube radius and $V_{max}$ characterizes the maximum velocity at the center of the tube. This nonlinear flow has a parabolic velocity profile, and in this case the distribution function cannot be obtained analytically. An expansion of the distribution for particles near the center of the flow has been calculated~\cite{van_de_Ven80}, making a change of variables to a translating system 
\begin{align}
x''&=x'-[x'(0)+v'(0)t'], \text{ and} \nonumber \\
y''&=y'-y'(0)\nonumber
\end{align}
 and then considering the dimensionless parameters 
\begin{align}
t&=Dt'/y_0, \nonumber \\
x&=x''/x_0, \nonumber \\
y&=y''/y_0, \text{ and} \nonumber\\
\sigma&=\gamma y_0^3/D, \nonumber 
\end{align}
 where $\sigma $ is a local P\'eclet number. In these variables, the convective diffusion equation becomes
\begin{equation}
\label{eq.pouseuille}
\partial P/\partial t = \nabla^{2}P+\sigma(2y+y^{2})\partial P/\partial x,
\end{equation}
with an approximate solution 
\begin{equation}
P=P_0+\sigma P_1+\sigma^2P_2+\cdots,
\end{equation}
 where the $P_i$'s are given by
\begin{equation}
P_i=p_i(x,y,t)\exp\left[-\frac{x^{2}+y^{2}}{4t}\right].
\end{equation}
The functions $p_i$ become more complex as we increase $i$: 
\begin{equation}
p_0=\frac{1}{8(\pi t)^\frac{3}{2}}, \qquad p_1=-\frac{x (t + 3y + y^{2})}{48 (\pi t)^{\frac{3}{2}}}, 
\end{equation}
  and for $p_2$, which is an even function of $x$, see Ref.~\cite{van_de_Ven80}. For this case also, no time-independent relations such as (\ref{eq.ft1}) can be found, as expected, since this is not a potential system. Instead, we obtain as an approximation for small $\sigma$
\begin{align}
&\frac{P(x,y,t)}{P(-x,y,t)}\approx 1+2\sigma \frac{P_1}{P_0}+\sigma^2 \frac{P_1^2}{P_0^2}\nonumber \\
&=1-\sigma \frac{x(t+3y+y^2)}{3} +\sigma^2\! \! \left[ \frac{x(t+3y+y^2)}{6}\right]^2 \! \! .
\end{align}
If $\sigma=0$, Eq.~(\ref{eq.pouseuille}) reduces to a diffusion equation and has a Gaussian solution; however, if $\sigma\neq 0$, then the solution will not be Gaussian, independent of whether $\sigma$ is small or not. Therefore, this illustrates a simple case of a system which may be very close to equilibrium but that displays non-Gaussian behavior of the probability distribution function.

\section{Conclusion}
\label{sec.conclusion}

We have studied model one-dimensional systems and a two-dimensional system composed of noninteracting Brownian particles and a liquid driven to a steady state by external forces in thermal equilibrium.   Although the flow conditions are steady, the colloidal particle distribution may not have a steady distribution as is the case of a colloidal suspension in an infinite region. Notwithstanding,  in potential systems, some time-independent relations can be found. 

We demonstrated that the same kind of time-independent relations  that appear in experiments with single molecules are present in Brownian motion in elongational flow. 
 By  showing that the origin of these relations is microscopic reversibility and the existence of a potential, we conclude that such relations are natural consequences of the thermodynamics of potential systems, even when considering nonequilibrium states.  

We also show that the Gaussian nature of a distribution is not necessarily related to equilibrium conditions. Even though the external driving force may be large, keeping the system in a far from equilibrium state, the distribution can be Gaussian nonetheless. We pointed out that the linear character of the flow is more important to Gaussianization than the equilibrium (or nonequilibrium) state of the system. This was seen in the fact that even though the external driving in Pouseuille flow can be small, the distribution function will not be Gaussian. On the other hand, in the linear cases, such as elongational flow or simple shear, the fluxes may be large and the distributions still be Gaussian.

In summary, we have shown the implications of the Gaussian nature of the noise in the stochastic behavior of non-equilibrium systems. Even  in  systems subjected to strong external forces, as in the stretching of single molecules, the process is carried out in a solution that serves as a thermal bath held at constant temperature and, therefore, there is no reason for the noise to be other than Gaussian. That, together with the potential nature of the system, leads naturally to the pervasive presence of time-independent relations in non-equilibrium systems.

\acknowledgments
This work was supported by the DGiCYT under grant no. FIS2005-01299, by CNPq and by CAPES.


\begin{thebibliography}{24}
\expandafter\ifx\csname natexlab\endcsname\relax\def\natexlab#1{#1}\fi
\expandafter\ifx\csname bibnamefont\endcsname\relax
  \def\bibnamefont#1{#1}\fi
\expandafter\ifx\csname bibfnamefont\endcsname\relax
  \def\bibfnamefont#1{#1}\fi
\expandafter\ifx\csname citenamefont\endcsname\relax
  \def\citenamefont#1{#1}\fi
\expandafter\ifx\csname url\endcsname\relax
  \def\url#1{\texttt{#1}}\fi
\expandafter\ifx\csname urlprefix\endcsname\relax\def\urlprefix{URL }\fi
\providecommand{\bibinfo}[2]{#2}
\providecommand{\eprint}[2][]{\url{#2}}

\bibitem[{\citenamefont{Smith et~al.}(1992)\citenamefont{Smith, Finzi, and
  Bustamante}}]{Smith92}
\bibinfo{author}{\bibfnamefont{S.~B.} \bibnamefont{Smith}},
  \bibinfo{author}{\bibfnamefont{L.}~\bibnamefont{Finzi}}, \bibnamefont{and}
  \bibinfo{author}{\bibfnamefont{C.}~\bibnamefont{Bustamante}},
  \bibinfo{journal}{Science} \textbf{\bibinfo{volume}{258}},
  \bibinfo{pages}{1122} (\bibinfo{year}{1992}).

\bibitem[{\citenamefont{Perkins et~al.}(1995)\citenamefont{Perkins, Smith,
  Larson, and Chu}}]{Perkins95}
\bibinfo{author}{\bibfnamefont{T.~T.} \bibnamefont{Perkins}},
  \bibinfo{author}{\bibfnamefont{D.~E.} \bibnamefont{Smith}},
  \bibinfo{author}{\bibfnamefont{R.~G.} \bibnamefont{Larson}},
  \bibnamefont{and} \bibinfo{author}{\bibfnamefont{S.}~\bibnamefont{Chu}},
  \bibinfo{journal}{Science} \textbf{\bibinfo{volume}{268}},
  \bibinfo{pages}{83} (\bibinfo{year}{1995}).

\bibitem[{\citenamefont{Wang et~al.}(2002)\citenamefont{Wang, Sevick, Mittag,
  Searles, and Evans}}]{Wang02}
\bibinfo{author}{\bibfnamefont{G.~M.} \bibnamefont{Wang}},
  \bibinfo{author}{\bibfnamefont{E.~M.} \bibnamefont{Sevick}},
  \bibinfo{author}{\bibfnamefont{E.}~\bibnamefont{Mittag}},
  \bibinfo{author}{\bibfnamefont{D.~J.} \bibnamefont{Searles}},
  \bibnamefont{and} \bibinfo{author}{\bibfnamefont{D.~J.} \bibnamefont{Evans}},
  \bibinfo{journal}{Phys. Rev. Lett.} \textbf{\bibinfo{volume}{89}},
  \bibinfo{pages}{050601} (\bibinfo{year}{2002}).

\bibitem[{\citenamefont{Carberry et~al.}(2004)\citenamefont{Carberry, Reid,
  Wang, Sevick, Searles, and Evans}}]{Carberry04}
\bibinfo{author}{\bibfnamefont{D.~M.} \bibnamefont{Carberry}},
  \bibinfo{author}{\bibfnamefont{J.~C.} \bibnamefont{Reid}},
  \bibinfo{author}{\bibfnamefont{G.~M.} \bibnamefont{Wang}},
  \bibinfo{author}{\bibfnamefont{E.~M.} \bibnamefont{Sevick}},
  \bibinfo{author}{\bibfnamefont{D.~J.} \bibnamefont{Searles}},
  \bibnamefont{and} \bibinfo{author}{\bibfnamefont{D.~J.} \bibnamefont{Evans}},
  \bibinfo{journal}{Phys. Rev. Lett.} \textbf{\bibinfo{volume}{92}},
  \bibinfo{pages}{140601} (\bibinfo{year}{2004}).

\bibitem[{\citenamefont{Shang et~al.}(2005)\citenamefont{Shang, Tong, and
  Xia}}]{Shang05}
\bibinfo{author}{\bibfnamefont{X.~D.} \bibnamefont{Shang}},
  \bibinfo{author}{\bibfnamefont{P.}~\bibnamefont{Tong}}, \bibnamefont{and}
  \bibinfo{author}{\bibfnamefont{K.~Q.} \bibnamefont{Xia}},
  \bibinfo{journal}{Phys. Rev. E} \textbf{\bibinfo{volume}{72}},
  \bibinfo{pages}{015301(R)} (\bibinfo{year}{2005}).

\bibitem[{\citenamefont{Douarche et~al.}(2006)\citenamefont{Douarche, Joubaud,
  Garnier, Petrosyan, and Ciliberto}}]{Douarche06}
\bibinfo{author}{\bibfnamefont{F.}~\bibnamefont{Douarche}},
  \bibinfo{author}{\bibfnamefont{S.}~\bibnamefont{Joubaud}},
  \bibinfo{author}{\bibfnamefont{N.~B.} \bibnamefont{Garnier}},
  \bibinfo{author}{\bibfnamefont{A.}~\bibnamefont{Petrosyan}},
  \bibnamefont{and}
  \bibinfo{author}{\bibfnamefont{S.}~\bibnamefont{Ciliberto}},
  \bibinfo{journal}{Phys. Rev. Lett.} \textbf{\bibinfo{volume}{97}},
  \bibinfo{pages}{140603} (\bibinfo{year}{2006}).

\bibitem[{\citenamefont{Astumian}(2006)}]{Astumian06}
\bibinfo{author}{\bibfnamefont{R.~D.} \bibnamefont{Astumian}},
  \bibinfo{journal}{Am. J. Phys.} \textbf{\bibinfo{volume}{74}},
  \bibinfo{pages}{683} (\bibinfo{year}{2006}).

\bibitem[{\citenamefont{Onsager and Machlup}(1953)}]{Onsager53}
\bibinfo{author}{\bibfnamefont{L.}~\bibnamefont{Onsager}} \bibnamefont{and}
  \bibinfo{author}{\bibfnamefont{S.}~\bibnamefont{Machlup}},
  \bibinfo{journal}{Phys. Rev.} \textbf{\bibinfo{volume}{91}},
  \bibinfo{pages}{1505} (\bibinfo{year}{1953}).

\bibitem[{\citenamefont{Wang et~al.}(2001)\citenamefont{Wang, Forbes, and
  Jin}}]{Wang01}
\bibinfo{author}{\bibfnamefont{K.}~\bibnamefont{Wang}},
  \bibinfo{author}{\bibfnamefont{J.~G.} \bibnamefont{Forbes}},
  \bibnamefont{and} \bibinfo{author}{\bibfnamefont{A.~J.} \bibnamefont{Jin}},
  \bibinfo{journal}{Prog. Biophys. Mol. Biol.} \textbf{\bibinfo{volume}{77}},
  \bibinfo{pages}{1} (\bibinfo{year}{2001}).

\bibitem[{\citenamefont{Purcell}(1977)}]{Purcell77}
\bibinfo{author}{\bibfnamefont{E.~M.} \bibnamefont{Purcell}},
  \bibinfo{journal}{Am. J. Phys.} \textbf{\bibinfo{volume}{45}},
  \bibinfo{pages}{3} (\bibinfo{year}{1977}).

\bibitem[{\citenamefont{Astumian}()}]{Astumian06a}
\bibinfo{author}{\bibfnamefont{R.~D.} \bibnamefont{Astumian}},
  \bibinfo{note}{cond-mat/0608352}.

\bibitem[{\citenamefont{Taniguchi and Cohen}()}]{Taniguchi06}
\bibinfo{author}{\bibfnamefont{T.}~\bibnamefont{Taniguchi}} \bibnamefont{and}
  \bibinfo{author}{\bibfnamefont{E.~G.~D.} \bibnamefont{Cohen}},
  \bibinfo{note}{cond-mat/0605548}.

\bibitem[{\citenamefont{Risken}(1989)}]{Risken89}
\bibinfo{author}{\bibfnamefont{H.}~\bibnamefont{Risken}},
  \emph{\bibinfo{title}{{The Fokker-Planck Equation}}}
  (\bibinfo{publisher}{Springer-Verlag}, \bibinfo{address}{Berlin},
  \bibinfo{year}{1989}).

\bibitem[{\citenamefont{Qian et~al.}(2002)\citenamefont{Qian, Qian, and
  Tang}}]{Qian02}
\bibinfo{author}{\bibfnamefont{H.}~\bibnamefont{Qian}},
  \bibinfo{author}{\bibfnamefont{M.}~\bibnamefont{Qian}}, \bibnamefont{and}
  \bibinfo{author}{\bibfnamefont{X.}~\bibnamefont{Tang}}, \bibinfo{journal}{J.
  Stat. Phys.} \textbf{\bibinfo{volume}{107}}, \bibinfo{pages}{1129}
  (\bibinfo{year}{2002}).

\bibitem[{\citenamefont{van Kampen}(1992)}]{vanKampen92}
\bibinfo{author}{\bibfnamefont{N.~G.} \bibnamefont{van Kampen}},
  \emph{\bibinfo{title}{Stochastic processes in physics and chemistry}}
  (\bibinfo{publisher}{North-Holland}, \bibinfo{address}{Amsterdam},
  \bibinfo{year}{1992}).

\bibitem[{\citenamefont{de~Groot and Mazur}(1984)}]{deGroot84}
\bibinfo{author}{\bibfnamefont{S.~R.} \bibnamefont{de~Groot}} \bibnamefont{and}
  \bibinfo{author}{\bibfnamefont{P.}~\bibnamefont{Mazur}},
  \emph{\bibinfo{title}{Non-equilibrium thermodynamics}}
  (\bibinfo{publisher}{Dover}, \bibinfo{address}{New-York},
  \bibinfo{year}{1984}).

\bibitem[{\citenamefont{Mazur and Bedeaux}(1991)}]{Mazur91}
\bibinfo{author}{\bibfnamefont{P.}~\bibnamefont{Mazur}} \bibnamefont{and}
  \bibinfo{author}{\bibfnamefont{D.}~\bibnamefont{Bedeaux}},
  \bibinfo{journal}{Physica A} \textbf{\bibinfo{volume}{173}},
  \bibinfo{pages}{155} (\bibinfo{year}{1991}).

\bibitem[{\citenamefont{{ter Haar} and Wergeland}(1960)}]{terHaar60}
\bibinfo{author}{\bibfnamefont{D.}~\bibnamefont{{ter Haar}}} \bibnamefont{and}
  \bibinfo{author}{\bibfnamefont{H.~N.~S.} \bibnamefont{Wergeland}},
  \emph{\bibinfo{title}{{Elements of Thermodynamics }}}
  (\bibinfo{publisher}{Addison-Wesley}, \bibinfo{year}{1960}).

\bibitem[{\citenamefont{Foister and {van de Ven}}(1980)}]{van_de_Ven80}
\bibinfo{author}{\bibfnamefont{R.~T.} \bibnamefont{Foister}} \bibnamefont{and}
  \bibinfo{author}{\bibfnamefont{T.~G.~M.} \bibnamefont{{van de Ven}}},
  \bibinfo{journal}{J. Fluid Mech.} \textbf{\bibinfo{volume}{96}},
  \bibinfo{pages}{105} (\bibinfo{year}{1980}).

\bibitem[{\citenamefont{Dufty and Rub\'{\i}}(1987)}]{PhysRevA.36.222}
\bibinfo{author}{\bibfnamefont{J.~W.} \bibnamefont{Dufty}} \bibnamefont{and}
  \bibinfo{author}{\bibfnamefont{J.~M.} \bibnamefont{Rub\'{\i}}},
  \bibinfo{journal}{Phys. Rev. A} \textbf{\bibinfo{volume}{36}},
  \bibinfo{pages}{222} (\bibinfo{year}{1987}).

\bibitem[{\citenamefont{Rub\'{\i} and P\'erez-Madrid}(1991)}]{Rubi91}
\bibinfo{author}{\bibfnamefont{J.~M.} \bibnamefont{Rub\'{\i}}}
  \bibnamefont{and}
  \bibinfo{author}{\bibfnamefont{A.}~\bibnamefont{P\'erez-Madrid}},
  \bibinfo{journal}{Phys. Rev. A} \textbf{\bibinfo{volume}{43}},
  \bibinfo{pages}{7040} (\bibinfo{year}{1991}).

\bibitem[{\citenamefont{Metropolis et~al.}(1953)\citenamefont{Metropolis,
  Rosenbluth, Rosenbluth, and Teller}}]{Metropolis53}
\bibinfo{author}{\bibfnamefont{N.}~\bibnamefont{Metropolis}},
  \bibinfo{author}{\bibfnamefont{A.~W.} \bibnamefont{Rosenbluth}},
  \bibinfo{author}{\bibfnamefont{M.~N.} \bibnamefont{Rosenbluth}},
  \bibnamefont{and} \bibinfo{author}{\bibfnamefont{A.~H.}
  \bibnamefont{Teller}}, \bibinfo{journal}{J. Chem. Phys.}
  \textbf{\bibinfo{volume}{21}}, \bibinfo{pages}{1087} (\bibinfo{year}{1953}).

\bibitem[{\citenamefont{Chandrasekhar}(1943)}]{Chandrasekhar43}
\bibinfo{author}{\bibfnamefont{S.}~\bibnamefont{Chandrasekhar}},
  \bibinfo{journal}{Rev. Mod. Phys.} \textbf{\bibinfo{volume}{15}},
  \bibinfo{pages}{1} (\bibinfo{year}{1943}).

\bibitem[{\citenamefont{Bochkov and Kuzovlev}(1981)}]{Bochkov81}
\bibinfo{author}{\bibfnamefont{G.~N.} \bibnamefont{Bochkov}} \bibnamefont{and}
  \bibinfo{author}{\bibfnamefont{Y.~E.} \bibnamefont{Kuzovlev}},
  \bibinfo{journal}{Physica A} \textbf{\bibinfo{volume}{106}},
  \bibinfo{pages}{443} (\bibinfo{year}{1981}).

\end{thebibliography}

\end{document}